\documentclass[default,iicol]{sn-jnl}

\usepackage{graphicx}%
\usepackage{multirow}%
\usepackage{amsmath,amssymb,amsfonts}%
\usepackage{amsthm}%
\usepackage{mathrsfs}%
\usepackage[title]{appendix}%
\usepackage{xcolor}%
\usepackage{textcomp}%
\usepackage{manyfoot}%
\usepackage{booktabs}%
\usepackage{algorithm}%
\usepackage{algorithmicx}%
\usepackage{algpseudocode}%
\usepackage{listings}%
\usepackage{makecell}
\usepackage{changepage}
\usepackage{array}
\usepackage{hyperref}
\usepackage[figuresright]{rotating}

\theoremstyle{thmstyleone}%
\theoremstyle{thmstyletwo}%

\theoremstyle{thmstylethree}%

\begin{document}

\title[Kellect: a Kernel-Based Efficient and Lossless Event Log Collector for Windows Security]{Kellect: a Kernel-Based Efficient and Lossless Event Log Collector for Windows Security}


\author{\fnm{Tieming} \sur{Chen}}\email{tmchen@zjut.edu.cn}

\author{\fnm{Qijie} \sur{Song}}\email{songqijie@zjut.edu.cn}

\author{\fnm{Xuebo} \sur{Qiu}}\email{qiuxuebo@zjut.edu.cn}
\author{\fnm{Tiantian} \sur{Zhu*}}\email{ttzhu@zjut.edu.cn}
\author{\fnm{Zhiling} \sur{Zhu}}\email{zhilingzhu@zjut.edu.cn}
\author{\fnm{Mingqi} \sur{Lv}}\email{mingqilv@zjut.edu.cn}

\affil{\orgdiv{College of Computer Science and Technology}, \orgname{Zhejiang University of Technology}, \orgaddress{\city{Hangzhou, Zhejiang}, \postcode{310012}, \country{China}}}

\abstract{
  Recently, APT attacks have frequently happened, which are increasingly complicated and more challenging for traditional security detection models. The system logs are fundamentally vital for cyber security analysis mainly due to their effective reconstruction ability of system behavior. Furthermore, research on dynamic detection and traceability for events with a long and hidden latency through system logs has been widely of concern. For Windows, ETW(events tracing for Windows) is a well-known built-in kernel-level logs collection framework. However, existing log collection tools built on ETW suffer from working shortages, including data loss, high overhead, and weak real-time performance. Therefore, It is still very difficult to directly apply ETW-based windows tools to analyze APT attack scenarios.
  
  To address these challenges, this paper proposes an efficient and lossless kernel log collector called Kellect,which has open sourced with project at www.kellect.org. Kellect is also developed based on the ETW but extremely efficient on log parsing and event analyzing. It takes extra CPU usage with only approximately 2\%-3\% and about 40MB memory consumption, by dynamically optimizing the number of cache and processing threads through a multi-level cache solution. By replacing the TDH library with a sliding pointer, Kellect significantly enhances analysis performance, achieving at least 9 times the efficiency of existing tools. Furthermore, Kellect improves compatibility with different OS versions. Additionally, Kellect enhances log semantics understanding by maintaining event mappings and application callstacks which provide more comprehensive characteristics for security event behavior analysis.
  
  With plenty of experiments, Kellect demonstrates its capability to achieve non-destructive, real-time and full collection of kernel log data generated from events with a comprehensive efficiency as 9 times greater than existing tools. Furthermore, log analysis module is implemented with lightweight but precise in Kellect, so it can be directly utilized for Windows security analysis such as APT detection and event tracing. As a killer illustration to show how Kellect can work for APT, full data logs have been collected as a dataset Kellect4APT, generated by implementing diversity TTPs from the latest ATT\&CK. To our best knowledge, it is the first open benchmark dataset representing for ATT\&CK technique-specific behaviors, which could be highly expected to improve more extensive research on APT study. 
}


\maketitle

\section{Introduction}
Endpoints are crucial as fundamental elements and integral components in constructing information networks. They are extensively deployed and utilized across a wide range of business functions. Endpoints are repositories for critical network data, making them highly susceptible to malicious attacks and data theft. In recent years, the prevalence of malicious threats targeting endpoints has notably escalated, along with the heightened risk of APT (Advanced Persistent Threat) attacks against advanced information security systems. Traditional methods of threat detection based on traffic analysis have demonstrated insufficiency in addressing the escalating need for robust security analysis and precise tracking. \cite{inam2023sok}.

To achieve integrated security governance in collaboration between cloud services and endpoints, Log-based analysis has become an essential requirement. Logs can be categorized into user level and kernel level. User-level logs are generated through human-computer interaction, capturing API calls and recording the behavior of program software. However, user-level data has shortcomings, such as unclear semantics and vulnerability to malicious tampering. Kernel-level logs record fine-grained underlying behavior by collecting system calls and related parameters. This reflects user and software behaviors better, and kernel-generated data is more reliable. Kernel-level logs have greater traceability and data source reliability than semantically inefficient network traffic. Therefore, many researchers have utilized kernel logs for cybersecurity research \cite{yang2020ratscope,yang2020ratscope,al2018studying,irshad2021trace,chen2021clarion,zeng2021watson}.

Windows is the dominant system in office environments with widespread adoption\cite{statcounter}. With the increasing development of attack technology,  It faces heightened vulnerability to network-based threats, including APTs, Trojan, and ransomware viruses. Kernel logs are widely used for Windows attack detection. Using kernel logs to establish causal relationships between system activities can effectively analyze and track attack behaviors\cite{ahmed2021peeler,chen2021clarion}. Despite Windows being relatively closed compared to Linux and lacking the same level of openness, Microsoft provides the ETW(events tracing for Windows) framework for kernel log tracking \cite{ntkernellogger}. ETW enables the collection of diverse log types from various configured providers, encompassing files, processes, networks, and more.

However, in the collection of kernel logs based on ETW, we found that there are serious data loss and performance problems in the existing technology. The specific problems are as follows:

\begin{itemize}
\item \textbf{Data Loss.} The loss of data presents a significant challenge in the context of log collection, particularly when system load increases. When substantial amounts of data are lost, it compromises the system activity's integrity, especially in the case of full kernel logs collection. This exacerbates the difficulty of preserving all pertinent information, further complicating the task. Data loss directly impacts the primary objective of collection, which is to establish and maintain the contextual causality between different entities to enable effective detection and traceability.

\item \textbf{Real-time.} The real-time feedback of behavior plays a critical role in log collection. In the context of APT attacks, attackersmay employ tactics such as maliciously restarting or forcefully disabling the log collector, which disrupts the data collection process. However, a significant challenge exists in many tools, as they exhibit a substantial time interval between the occurrence of a behavior and the recording of the corresponding log \cite{paccagnella2020logging}. As the tool continues to run, this time interval progressively lengthens, failing to capture the most recent attack behaviors. Moreover, some behaviors necessitate offline analysis, further impeding the ability to meet real-time requirements in attack scenarios.

\item \textbf{Runtime Overhead.} Deploying data collection tools on endpoint devices should not disrupt daily office operations. However, existing collectors impose significant overhead, which can further escalate with prolonged running time. To ensure uninterrupted office productivity, it is essential to strike a balance between efficient data collection and minimal performance impact to mitigate the adverse effects on the endpoints' performance.

\item \textbf{Semantic Missing.} The presentation of log information presents challenges in terms of user-friendliness. Logs encompass diverse heterogeneous information, such as files, processes, threads, and networks. Analyzing such granular data using conventional methods can be difficult. Additionally, there are instances where certain events do not provide direct access to semantic information from the ETW, which limits the availability of detailed event attributes. Despite efforts in certain studies to transform logs into provenance graphs, the issue of missing specific event attribute information persists, impeding comprehensive analysis and comprehension.

\item \textbf{Large Data Volume.} Large amount of data will be generated. The raw data collected on a single client computer can reach 5GB per day, and the raw data collected by an ordinary office network with thousands of people can even be as high as 15TB per day. However, the average latency time of an APT attack from initial intrusion to impact is 83 days, and the logs generated by an attack in an enterprise can reach the PB level \cite{xu2016high}.

\end{itemize}

Based on the above problems, we propose a lossless, real-time, efficient Windows kernel log collector called Kellect. ETW is a built-in log framework of Windows, so it will not cause the system's crash, and will not occupy too much system resources. We use the ETW to track the logs on the kernel side and carry them out according to different providers provided by ETW of log parsing. Kellect provides almost all provider parsing templates. Although there may be some differences in different OS versions, we have made tremendous development efforts on tool adaptivity so that Kellect can run in all distribution systems of Windows 7 and its subsequent versions.

Below we summarize our approach and contributions:
\begin{itemize}
\item  Our R\&D motivation comes from the embarrassments of existing tools. We found that the widely used Windows log collector have the following problems:
\begin{itemize}
  \item[-] High overhead, especially as the collecting time increases, the overhead increases exponentially.
  \item[-] Severe data loss, especially in high-load environments where the loss rate increases.
  \item[-] Behavior recording intervals are long, and behavior cannot be fully recorded before the process is shut down.
  \item[-] The semantics are missing. The tools collect only part of the logs of kernel, and cannot fully restore the attributes and specific behaviors of the object.
\end{itemize}
 \item We designed and implemented an efficient, non-destructive and real-time Windows log collector, including:
 \begin{itemize}
  \item[-] For large volume of data to guarantee full performance without data loss, we have introduced a multi-level caching mechanism to balance performance and parsing speed.
  \item[-] For data parsing speed-up, we improved the parsing method of the data frame, adopted sliding pointer parsing, which increased the parsing speed by 9 times, and abandoned the TDH\cite{retrievingeventdata} library provided by Windows.
  \item[-] For data defect and semantic loss, we construct entity mapping relationship to realize the completion of defect semantics. We also leverage call stack information to enhance the semantics of event behavior.
\end{itemize}
\item  We evaluated the implementation with reasonable experiment study, and the results showed that Kellect has the following advantages:
 \begin{itemize}
  \item[-] It can effectively avoid data loss when collecting data in a high-load environment.
  \item[-]In a common Windows with Kellect installed, the CPU consumption is stable at about 2\%-3\%, and the memory consumption is stable at about 40MB.
  \item[-]The time interval window can reach 6 times that of existing tools , and the interval from data generation to recording is within 1 second, which can be regarded as almost real-time.
\end{itemize}
\item  To show the power to improve APT research, a first opened benchmark dataset called Kellect4APT is finally introduced:\cite{attck}:
\begin{itemize}
  \item[-] The data is generated by implementing diversity TTPs from the latest ATT\&CK.
  \item[-] Kellect4APT opens for both academic and industry in-deep study on APT technique-specific behavior detection and intelligence analysis. \cite{li2021threat, CHEN2023103485}.
\end{itemize}
\end{itemize}

\section{Motivation Experiments}

Due to the closed nature of the Windows kernel, developers are unable to target the kernel for richer extensions and applications, so the choice of data sources for Windows audit log collection is relatively limited. Compared with audit log collectors on Linux(e.g. sysdig\cite{sysdig}, auditd\cite{linuxaudit} and eBPF\cite{ebpftrace}), the audit log collector of Windows only relies on the ETW provided by Microsoft. In addition, due to the perfect security rules and closed system of Windows itself, many researchers and attackers are more inclined to choose the Linux, but there are still serious challenges in the security of Windows.

Considering our goal of obtaining complete and real-time Windows kernel logs, the applications and systems that integrate ETW and other data sources do not provide the required data collection interface and are incongruent with our objective. Although the toolkits offer numerous user-friendly API interfaces, it remains too low-level to form an upper-level security tool with richer functionalities. Regarding the current collectors, Spade\cite{SPADE} and Conan\cite{xiong2020conan} are not suitable as references, and Wtrace\cite{wtrace} has incomplete data collection sources. Both SilkETW\cite{SilkETW} and Sealighter\cite{Sealighter} are open source collection tools available on GitHub, with numerous stars. Sealighter outperforms silkETW according to comparative analysis [27]. So in this paper ,we largely focus on the best performance of sealighter to compete.

We aim to configure Sealighter to achieve full kernel log collection to the best of our ability. We follow the documentation provided by each system and adhered to best practice recommendations from online resources. We exhaustively explore available performance-related configuration parameters and selected the settings that yielded the best performance. Our sole guiding principle is the comprehensive collection of source information for all processes. Specifically, we configure Sealighter to log events related to coarse-grained sources, including operations among file, process and network. The experimental events describe in this paper are based on this configuration. For detailed information regarding the specific data types collected in our study, please consult Appendix \ref{sec:EventsCollecting}.

\textbf{Experimental Environment Settings.} The experimental host environment consists of a Windows 10 Professional Workstation Edition equipped with an Intel i5-10400 processor (2.9GHz, 6 cores and 32GB of memory). The OS version is 21H2. To ensure consistent operating environments, we employe VirtualBox virtualization, ensuring identical configurations for both systems without mutual interference. The virtual machine is allocated 3 cores, 16GB of memory, and a 120GB SSD. All experiments conducted in this paper refer to this experimental environment setting.

\textbf{Benchmark Selecting.} To ensure event consistency, we select two benchmarks as evaluation criteria. Firstly, pcopy, a script tool for file reading and writing in the Windows, is utilized. We design a script that utilized multithreading for large file operations, simulating file I/O load by simulating the copying and moving of large files using pcopy. Additionally, JMeter\cite{halili2008apache} is employed as a high-concurrency testing tool to assess the system's load capacity.

\begin{figure}[!htp] 
  \centering 
  \includegraphics[width=3.2in]{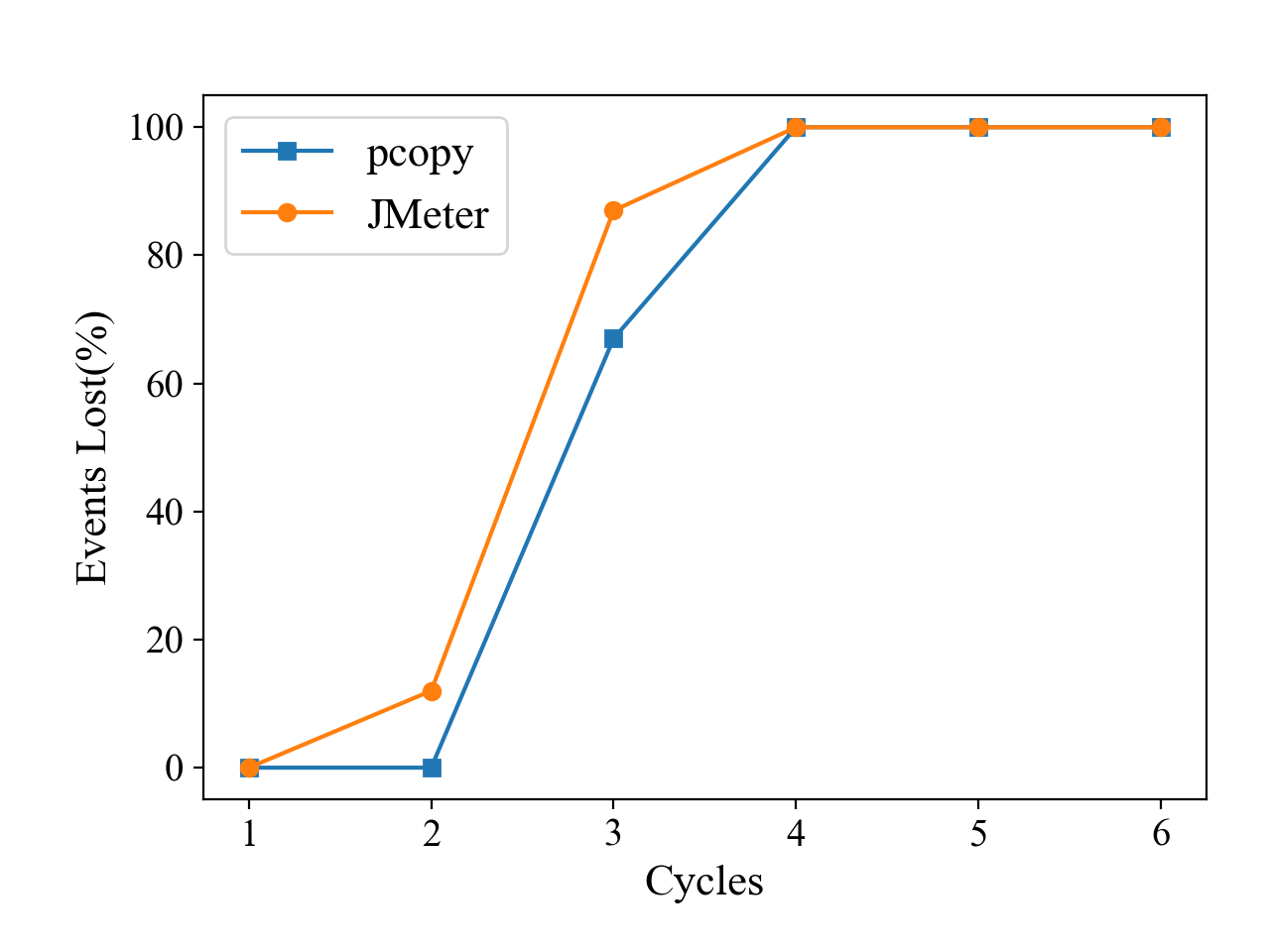}
  \caption{Eventlost of sealighter} 
  \label{Fig.main1} 
  \vspace{-1em}
\end{figure}

\textbf{Data Loss Comparison Setting.} We employe two test benchmarks to facilitate the collection and analysis of logs, and establish a method for evaluating data loss. The pcopy  is utilized to conduct paired file reading and writing, while the JMeter is employed to simulate the starting and closing of processes and threads in pairs. Specifically, we configure the tests to involve transferring 10GB of file data and creating 100000 threads within a 60-second timeframe, ensuring completion of the data request. We repeate this cycle six times. The experimental results are presented in Figure 1. Notably, in the second cycle, Sealighter begin to loss data, and in the third cycle, loss of filesystem logs is observed. The JMeter tool generates a substantial number of threads upon startup, resulting in a significant influx of thread logs that Sealighter struggles to process in a timely manner, consequently leading to data loss.

\textbf{Real-time Comparison Setting.} For real-time evaluation, we randomly selected a point during runtime to terminate the benchmark script, recording the current number of logs. The system continued parsing and recording the logs that were stuck in the cache. As the CPU usage decreased, signifying the completion of the phase, we recorded the number of logs once again. We calculated the difference between the two quantities of logs and repeated this process five times to obtain the results. Our analysis revealed that there were still 109,863 unprocessed data entries in the JMeter test and 87,643 unprocessed data entries in the pcopy test. Furthermore, we observed that Sealighter's data processing reached a plateau over time, as illustrated in Figure 2, indicating that a processing threshold had been reached. However, considering the duration, there remained a substantial volume of raw data for just a few seconds, presenting an opportunity for malicious actors to cover their activities.

\begin{figure}[!htp] 
  \centering 
  \includegraphics[width=3.2in]{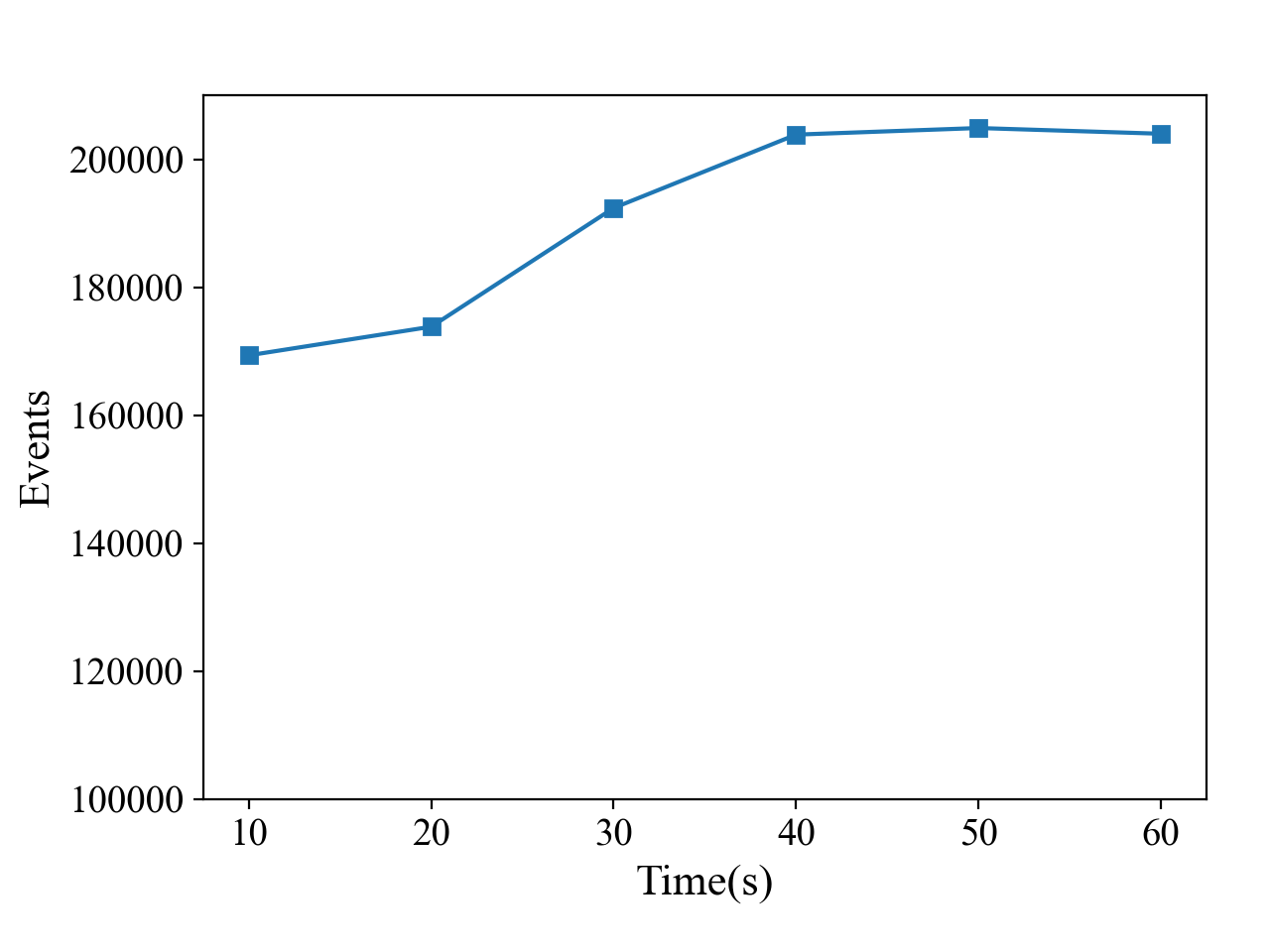}
  \caption{number of events over Time} 
  \label{Fig.main1} 
  \vspace{-1em}
\end{figure}

\textbf{Runtime Overhead Comparison Setting.} In terms of performance consumption, Sealighter's performance can be considered suboptimal. As depicted in Table 1, the tool exhibits a consistent CPU usage of approximately 40\%, and its memory consumption increases progressively as the duration of operation extends. This indicates that the parsing process encounters a bottleneck, restricting the tool's capability to store log data solely in memory. Consequently, when dealing with APT attacks that may persist for several days, Sealighter proves inadequate for handling the prolonged duration and volume of log data associated with such attacks.

\begin{table}[h]
  \caption{Overhead of Sealighter}
  \label{tab:freq}
  \begin{tabular}{p{2.2cm}p{2.2cm}p{2.2cm}}
    \toprule
    & CPU & Memory \\
    \midrule
    pcopy & 37.65\% & 64.08MB \\
    JMeter & 41.24\% & 73.54MB \\
  \bottomrule
\end{tabular} 
\end{table}

The analysis reveals a discernible gap in achieving comprehensive collection of Windows kernel logs, with the absence of an efficient tool to address this challenge. In light of these existing limitations, this paper presents a novel solution: a comprehensive, real-time, and lossless Windows kernel log collector.

\section{System design}

Aiming at the gap in the second section, we clarifie the goal of the Windows kernel log collector, and put forward our scheme based on the goal. The ETW of Windows is a typical producer-consumer model\cite{ntkernellogger}. In this paper, we believe that the ETW provided by Windows itself does not have problems such as missing data, and can ensure that the data provided for consumption is real-time and has complete semantics. It will not affect performance due to different data volumes caused by different providers, and assume that it is absolutely safe.

\subsection{Goal and Benchmark Setting}

This paper aims to develop an efficient log collection technique for the Windows kernel that operates seamlessly alongside normal workload activities. We have set three goals:  

\textbf{G1: Improve the Collection Configuration to Avoid Data Loss}.

\textbf{G2: Improve Data Consumption Efficiency to Ensure Real-time Collection}. 

\textbf{G3: Obtain Full Kernel Logs with Complete Semantics}. 

It is recognized that singular-dimensional kernel data fails to fully capture the intentions of attackers from a security perspective. Therefore, acquiring the entire set of kernel logs enables effective analysis of malicious behaviors throughout the entire attack process, facilitating comprehensive traceability analysis and preventive measures. However, it is acknowledged that acquiring and analyzing such extensive logs will increase data volume and impact overall system performance. Thus, a balance needs to be struck between comprehensive data parsing and performance consumption.

Real-time data collecting is also a key focus. Attackers may attempt to sabotage the log collecting process to prevent their actions from being exposed. Therefore, the collector need to swiftly and accurately capture, analyze, and record the behaviors of system users (including attackers) to minimize time discrepancies between data generated by the ETW and the final analysis recording. This ensures the comprehensive recording of system behaviors prior to the termination of the collecting process.

Furthermore, the paper emphasizes the semantic integrity of the audit logs. Log objects encompass various attributes of different entities, such as processes, files, and networks. Analyzing the relationships between behaviors alone may not accurately define specific attack operations, particularly in the case of complex attack behaviors. Hence, it is crucial to preserve the semantics of log objects as comprehensively as possible, including details such as commandlines, file paths, target addresses, data sizes in network communications, and even system calls. This comprehensive preservation guarantees the thorough reconstruction of attack behaviors, even at the cost of additional storage space.

It is important to note that the ETW serves as the built-in kernel log collection framework. When specifying the need to collect kernel logs, data sources are defined through configuration files and filtering mechanisms. As a result, modifications to the log source cannot be made when the data volume becomes excessively large. While various research works have explored time-space compression techniques for generated logs through semantic analysis and encoding, these compression operations occur after parsing and are beyond the scope of this paper's detailed discussion.

To evaluate the proposed tool, six benchmarks shown in Table 2 were selected, and corresponding scripts were developed for their execution. The table indicates the usage of pcopy and iperf\cite{tirumala1999iperf} for file system and network load peak tests, respectively. JMeter was utilized for CPU load testing by creating multiple threads. Additionally, compression software RAR, file indexing software Everything, and download tool IDM were employed. These software applications extensively invoked various APIs to evaluate the throughput load on the file system, CPU, and network, respectively.

\begin{table}[h]
  \caption{Benchmarks}
  \label{tab:Benchmarks}
  \begin{tabular}{p{2.0cm}p{4.8cm}}
    \toprule
    benchmarks & description\\
    \midrule
    pcopy & Script to test file read and write  \\
    iperf &  Network testing tools \\
    JMeter & Concurrency Test Tool  \\
    rar & A compression software to  test FileIO  \\
    everything & Indexing tool to testing file searches \\
    IDM & A downloader to test network \\
  \bottomrule
\end{tabular} 
\end{table}

\subsection{Overview of ETW}

ETW(Event Tracing for Windows) is a comprehensive and powerful framework designed by Microsoft for events tracing and analysis in the Windows. It provides the ability to capture and analyze various events, enabling deep understanding of the system, performance evaluation and problem diagnosis.

ETW operates through the collaboration of event providers, event consumers, and event sessions. Microsoft provides 1007 providers that cover various aspects of system behavior, such as application activity, hardware events, and kernel-level operations. Event consumers include built-in Windows components, third-party tools, and custom applications that collect and process these events for analysis and monitoring purposes.

\subsection{Architecture}

In this section, we present the overall design and introduction of the Kellect. Kellect is mainly divided into three main modules: initialization module, collecting module and processing module as Figure 3 shows.

\begin{figure}[!htp] 
  \centering 
  \includegraphics[width=3.0in]{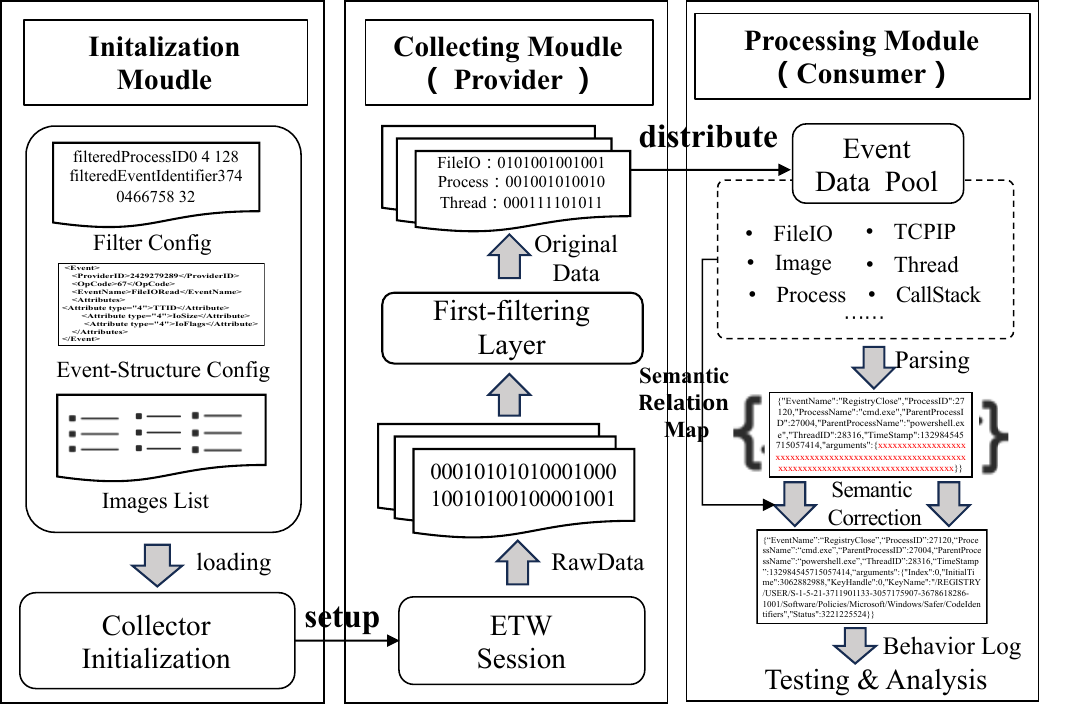}
  \caption{the Architecture of Kellect} 
  \label{Fig.main1} 
  \vspace{-1em}
\end{figure}

\textbf{The initialization module} is used to configure the collector before it runs. This module is responsible for loading the filtering process, event structure and mirror list, so as to prepare for the startup of ETW. In Windows, some processes are assigned fixed process IDs by the system. For example, 4 means system, and 148 means register. These processes are regarded as system default processes, and their IDs will not change with the start and stop of the system. Therefore, we consider them trustworthy and filter processes during collection to reduce unnecessary log processing overhead.

\textbf{The collecting module} is developed based on ETW. In the ETW, data is generated by producers and consumed by applications. The module obtains the raw data stream through the callback function. The original data stream contains a large amount of kernel operation information, most of which has missing information and incomplete semantics. This module performs as a layer of filtering on these data in combination with the previously loaded filtering process to remove data that cannot be parsed. In this module, we save the raw data stream that has been filtered by one layer in the raw data pool, and consume data through multi-threading to achieve the best performance.

\textbf{The processing module} is the core module of the collector, which semantically parsing the original data. According to the loaded event structure, Microsoft defines the kernel log as 13 different events. Different events have different opcodes and attributes. Therefore, the logs need to be distributed according to the event name and sent to the specified processing method for analysis. The original data is semantically corrected during parsing to supplement missing information or correct incomplete semantics.

\subsection{S1: Session Setting to Avoid Data Loss}

This section details the solution for G1 through session settings.

Unlike Linux, the log's structure cannot be modified at the kernel level because of not open-source in Windows. However, we found that the session configuration largely affects the degree of data loss. A session is a trace in ETW used to enable log collection, which defines the scope and configuration of event collection. During the session configuration process, we found that the buffer directly impacts the event loss rate.

ETW's event buffer adopts a FIFO strategy, new event data will be written to the buffer's current position, and then the reader can read evens from the beginning of the buffer. When a buffer page is full, new events will be written to the next available buffer page, and the cycle will continue. 

Several parameters in the session, such as Wnode.BufferSize, NumberOfBuffers, FreeBuffers, BufferSize, are buffer-related which will affect efficiency of collecting:

\begin{itemize}

\item[-] \textit{FreeBuffers} represents the number of free buffers currently available, which decreases as the buffer is used and increases as events in the buffer are processed and released. 

\item[-] \textit{NumberOfBuffers} is used to specify the number of buffers in the session. A larger value can increase the concurrency and capacity of the session to handle high-frequency event generation better.

\item[-] \textit{Wnode.BufferSize} is used to specify the size of each buffer, indicating the amount of event data each buffer can hold, in bytes. A larger BufferSize can accommodate more event data, reducing the possibility of buffer overflow and event loss.

\item[-] \textit{BufferSize} indicates the total size of the buffer used by the entire session, in units of pages, calculated by multiplying Wnode.BufferSize by NumberOfBuffers. It represents the total capacity of buffer resources allocated by the session and indirectly affects the efficiency of ETW processing.

\end{itemize}

By studying related literature \cite{ntkernellogger}, we found that FreeBuffers and NumberOfBuffers jointly determine the number of trace buffers currently available. Larger NumberOfBuffers values can increase session concurrency and capacity to handle high-frequency event generation better. In the actual development process, we found that the system will dynamically adjust the NumberOfBuffers and FreeBuffers parameters according to actual needs to achieve the best performance. Therefore, we use the default configuration for this parameter in this paper.

BufferSize is the total size of buffers used by the entire ETW session in pages. Its size is determined by the product of Node.Buffer Size and Number Of Buffers. When we set BufferSize, NumberOfBuffers will be dynamically adjusted. We limit the buffersize to 32-1024KB, and conduct data loss experiments on different benchmarks. The eventlost is provided in the session configuration type EVENT\_TRACE\_PROPERTIES of ETW to obtain the loss of events. The result is shown in the figure 4. As the BufferSize increases, the amount of eventloss at intervals gradually decreases. IDM and everything no longer lose events at 128KB. JMeter loses an average of 13 events, and several other benchmark tests are no longer lost when it reaches 256KB. But 2763 events are lost each time When it reaches 32KB. We think that in a high-concurrency environment, the number of events generated is much higher than other benchmarks, so a larger buffer size is required. After subsequent event stripping analysis, this is indeed the case.
\begin{figure}[!htp] 
  \centering 
  \includegraphics[width=3.2in]{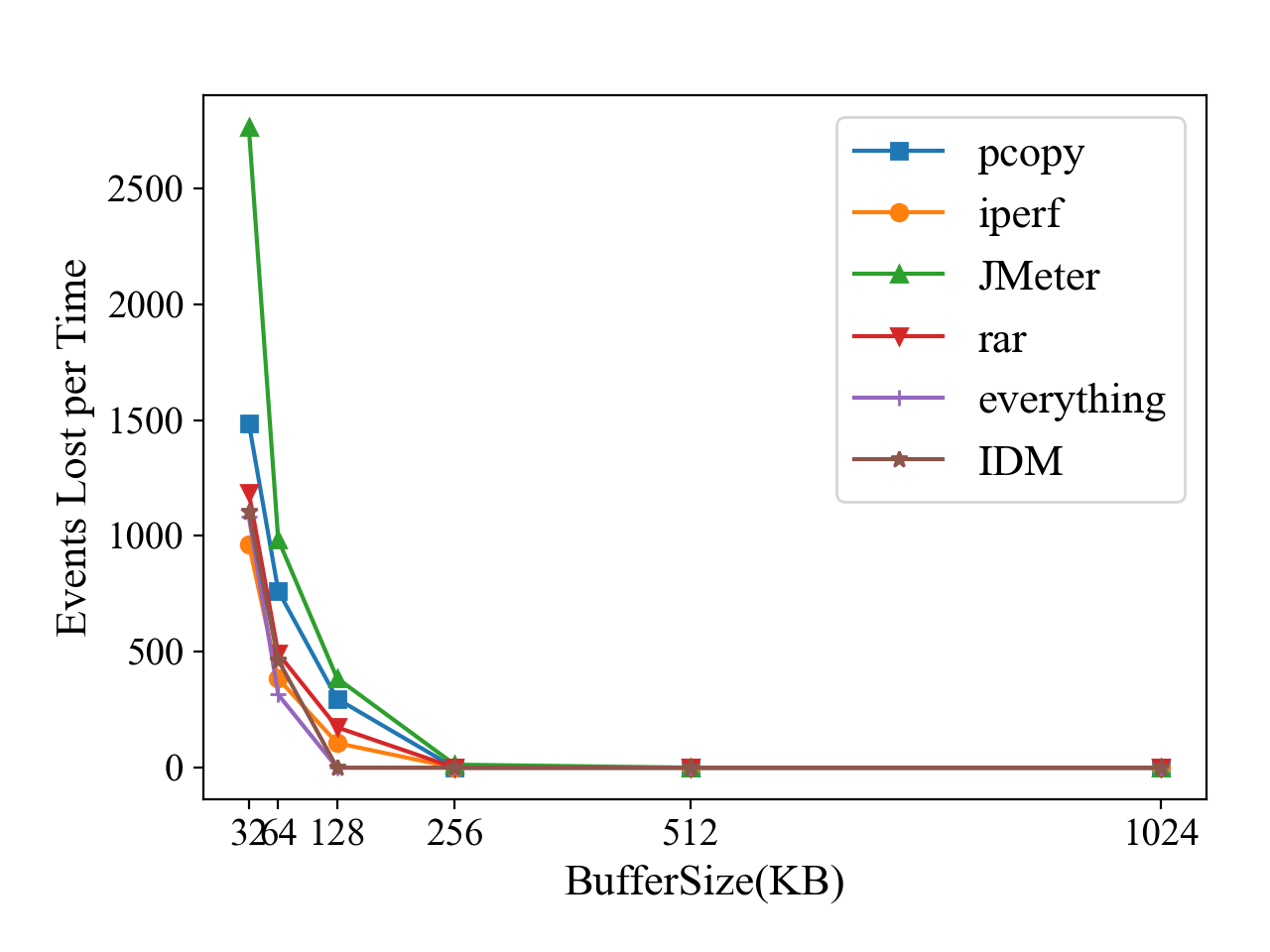}
  \caption{Event Lost over BufferSize Increasing} 
  \label{Fig.main1} 
  \vspace{-1em}
\end{figure}

We also test the overhead of the benchmark under different buffersizes as table 3 shows. We read the overhead of 5s before and after the loss and repeated 5 times to get the average value. As shown in the figure, we found that the overhead of memory and CPU under different benchmarks is almost the same, and has nothing to do with whether events are lost. By analyzing the caching mechanism of ETW, we found the parameters representing the buffer in the session. It can be believed that a perfect cache dynamic adjustment mechanism has been implemented inside the session. When an event is lost, RealTimeBuffersLost and LogBuffersLost are still 0, indicating that the buffer will not be affected by the loss of the event. The buffer is directly related to memory consumption.

The above analysis shows that the buffersize can be selected between 512KB and 1024KB, so we set the minimumbuffers to 512KB and the maximumbuffers to 1024KB. In this experiment, we set the buffersize of Kellect to 1024KB.

\begin{table}[h]
  \caption{Overhead of Different Buffersize}
  \label{tab:freq}
  \begin{tabular}{p{2.2cm}p{2.2cm}p{2.2cm}}
    \toprule
     & CPU & memory\\
    \midrule
    pcopy &0.2	&32.8  \\
    iperf & 0.1	&33.4 \\
    JMeter & 0.3&	40.1 \\
    rar &  0.1&	34.5\\
    everything & 0.1	&30.7 \\
    IDM &  0.2	&31.2\\
  \bottomrule
\end{tabular} 
\end{table}

\subsection{S2: Two Consumption Strategies to Achieve Real-time Data Collection}

In this section, we design two mechanisms to increase the consumption rate of events to achieve G2: \textbf{\textit{sliding pointer}} and \textbf{\textit{multi-level cache}}.

\textbf{Part 1: Use the sliding pointer to get event properties faster.}

 We use a sliding pointer to analyze the specific semantic of event attributes. Microsoft provides the TDH library\cite{retrievingeventdata} as an analysis tool for NT kernel log, but this tool needs to call multiple functions to obtain values, and the performance overhead is high, which cannot meet the high-performance requirements of kernel collection. Therefore, we propose an event parsing method based on the sliding pointer. After analyzing the original data, it is found that Windows sets the original event data as a fixed-length data frame and intercepts the number of bytes in different positions in the data frame according to different attribute offsets to parse the data correctly. 

Therefore, we preprocess the attributes contained in the 13 event types to form the event structure file in the initialization module. At this time, the collector has saved several groups of Event structure mapping relationship in the memory, the key-value pair in the pair is the attribute and its offset. The original data stream of the RegistryCreate event is 34-bit binary data. According to the attribute offset set, the first 8-bit binary data obtained first is parsed into PULONG type data, and the corresponding InitialTime attribute is obtained, which is the time value of the registry operation. Then the data stream is shifted to the left by the corresponding number of parsing digits, and the next attribute value is continued to be parsed based on the attribute offset set. Next, parse the 4-digit binary data into a PUSHORT type variable, and its value corresponds to the Status attribute, which is the result status of the registry operation. Loop through until the binary data stream is traversed, and finally get a complete event instance.

\begin{figure}[!htp] 
  \centering 
  \includegraphics[width=3in]{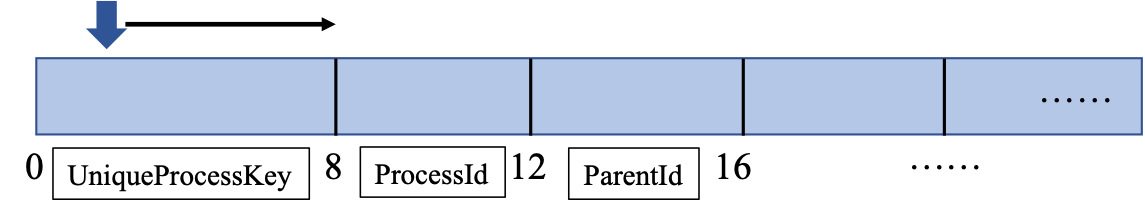}
  \caption{Analysis of Sliding Pointer} 
  \label{Fig.main1} 
  \vspace{-1em}
\end{figure}

The table 4 shows that parsing directly through the event structure greatly increases the efficiency of parsing, which is nearly 9 times faster than tdh parsing. At the same time, through the offset setting of different system events, the corresponding attributes can be obtained conveniently, without being limited by the TDH version, and more universal.

\begin{table}[h]
  \caption{Events Processed per Second}
  \label{tab:freq}
  \begin{tabular}{p{2.2cm}p{3.2cm}}
    \toprule
     & Parsed event per second \\
    \midrule
    TDH & 81094 \\
    Kellect &  749052 \\
  \bottomrule
\end{tabular} 
\end{table}

\textbf{Part 2: Use multi-level caching to speed up event consumption.} 

In the preceding section, we provide a comprehensive explanation of the buffer configuration for ETW's session. By appropriately setting the buffer size, ETW can effectively prevent data loss during the generation of kernel logs. However, as previously mentioned, ETW operates on a typical producer-consumer model, where the provider generates a substantial volume of logs that are then stored in the buffer. If the consumption of these logs does not occur in a timely manner, the buffer may still face the risk of overflowing, resulting in data loss. Our data analysis revealed that handling the massive influx of messages in the buffer using a single-threaded approach for data fetching and parsing is inadequate. Therefore, we developed a multi-level buffering solution to address the challenge of excessive buffer data.

In the solution, a buffer on the consumer side is used to receive messages. Given the exponential growth in the generation of full kernel logs, we implement multiple buffers and employed multi-threaded data consumption to ensure swift buffer clearance. However, since the number of kernel logs can fluctuate with system operations, it is crucial to determine the appropriate number of multi-level buffers and threads to prevent excessive consumption and resource waste.

The performance of multi-level cache consumption is influenced by several parameters, including the number of cache pools, the number of message parsing threads, the size of a single buffer, and the rate at which single-threaded data is parsed. The efficiency of single-threaded data analysis is enhanced by employing sliding window analysis, resulting in a tenfold increase in analysis efficiency. The size of the ETW cache is determined by the product of Wnode.BufferSize and NumberOfBuffers. Thus, it is necessary to ensure that the total amount of the product exceeds the size of the ETW cache. To investigate the impact of the number of threads on event loss, we initially set the multi-level buffer pool to 1/3 of the ETW session cache, enabling the rapid observation of data loss effects, and start the threads after a period of time as 10s. By subjecting the system to full load through the execution of JMeter, we obtained experimental results depicted in Figure 6. The figure demonstrates that when the number of threads is set to 4, data loss gradually decreases, achieving a balance with the producer. Consequently, we determine that setting the thread count to 4 is optimal.

Following the determination of the number of threads, we proceed to evaluate the size and number of buffer pools. While ensuring that the minimum total cache requirement is met, we base our judgment on CPU usage and incrementally increase the number of buffer pools. Figure 7 illustrates the results obtained from this evaluation. Notably, regardless of the increase in buffer pool size, CPU usage remained stable within the range of 30\% to 40\%, indicating that the size of the buffer pool does not significantly impact system overhead. It is important to note that Buffersize refers to the size of wnode.buffersize, and the number of caches is dynamically adjusted based on the availability of FreeBuffers.

\begin{figure}[!htp] 
  \centering 
  \includegraphics[width=3.2in]{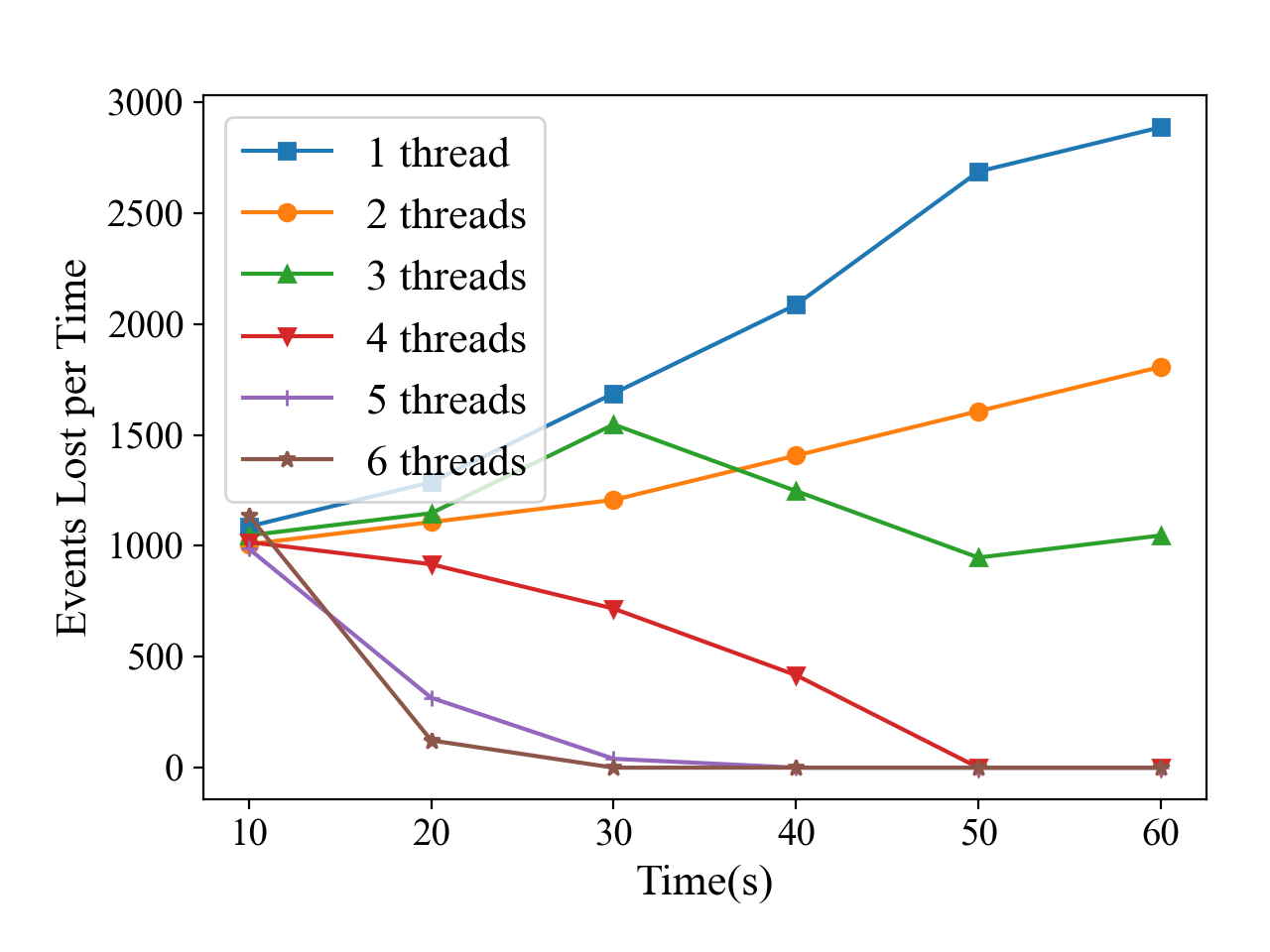}
  \caption{Event Lost with Different Number of Threads} 
  \label{Fig.main1} 
  \vspace{-1em}
\end{figure}

\begin{figure}[!htp] 
  \centering 
  \includegraphics[width=3.2in]{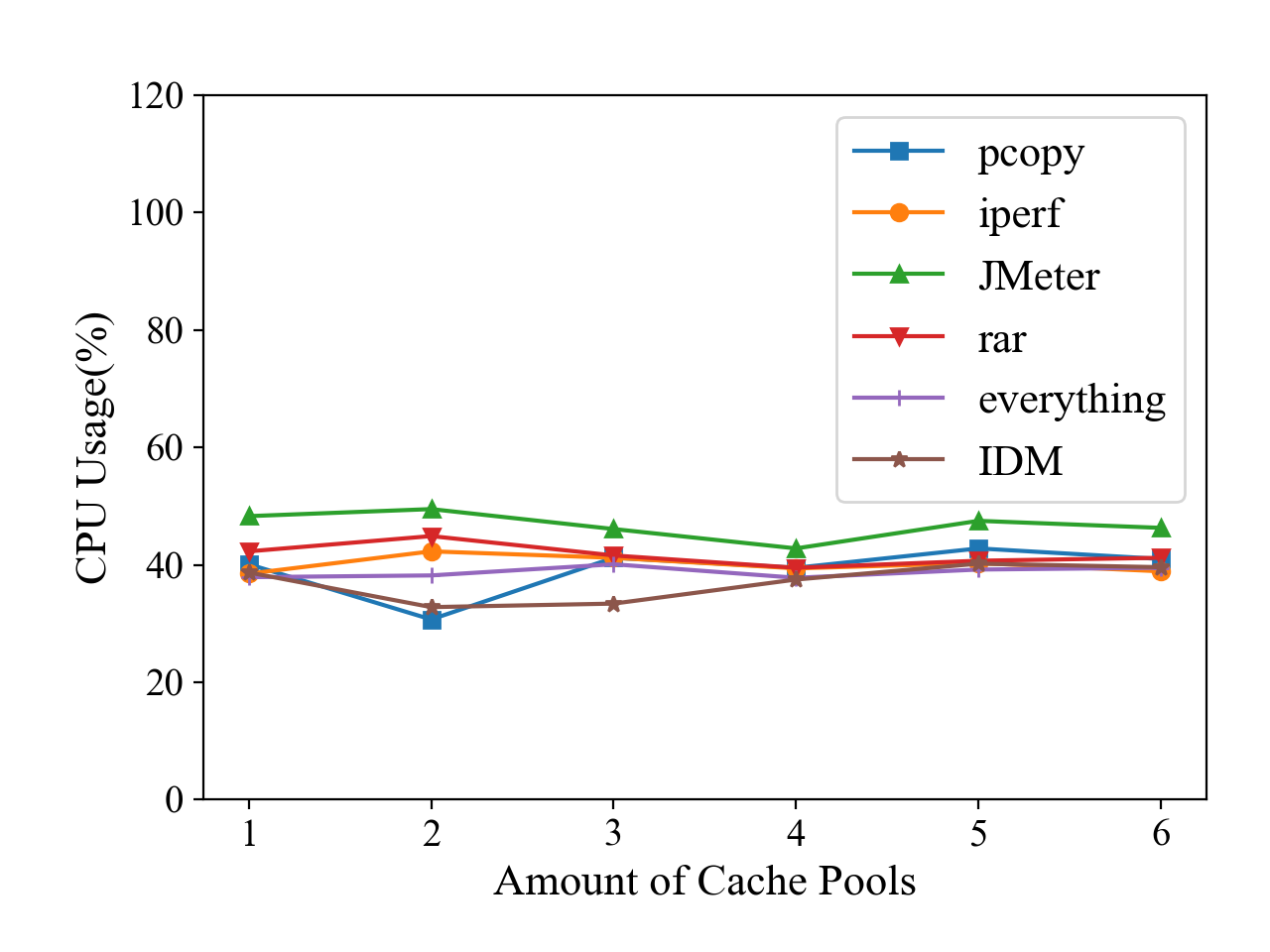}
  \caption{CPU usage over the amount of cache pools} 
  \label{Fig.main1} 
  \vspace{-1em}
\end{figure}

\subsection{S3: Associated Attributes to Modify Semantics}

In this section, we use relationship mapping tables and callstacks in the system to modify semantics to achieve G3.

\textbf{Part 1: Maintaining Relationship Mapping Tables to Fix Semantics.} 

After parsing and processing events, it is common to encounter situations where the semantics of certain attributes remain unclear. For instance, in the context of FileIO events, the absence of the FileName property in certain events, such as the FileRead, necessitates semantic augmentation to include the missing file name information. Additionally, the default value of 0xffffffff for the thread id fails to accurately represent the corresponding thread object for native TCP/IP events.

To address these semantic ambiguities, Kellect maintains 18 relational mapping tables based on the information collected during different processing stages. Table 5 exemplifies the coverage of all possible default semantic situations, which are pre-stored and parsed. For instance, the mapping table, derived from the relationship between process numbers and thread numbers in the CSwitch event, facilitates semantic correction for the default case of TCP/IP thread id. This allows the retrieval of the actual thread and host process associated with the network event.

\begin{sidewaystable*}[htp]
  \begin{adjustwidth}{-1cm}{}
  \caption{Mapping Lists to Parsing Events}
  \begin{center}
      \setlength{\tabcolsep}{2mm}{
  \begin{tabular}{p{0.8cm}p{5.2cm}p{8.2cm}p{4.2cm}}
  \hline
  \textbf{Index}&\textbf{Mapping Relation}&\textbf{Effection}&\textbf{Where and When Generated} \\
  \hline
  1  & ProcessId2ThreadId &Conversion from ProcessId to ThreadId & CSwitch in Thread Event \\
  \hline
  2	&	ThreadId2ProcessId&	Conversion from ThreadId to ProcessID &Thread\\
  \hline
  3 &	FileObject2Name&	Conversion from Fileobject to Filename&	\multirow{2}*{FileIO} \\
  \cline{1-3}
   4 & FileKey2Name	&Conversion from FileKey to FileName &\\
  \hline
  5	& ProcessID2ModuleAddressPair	& Conversion from Processid to Moduleaddresspair	& CallStack\\
  \hline
  6& 	ProcessName2Id& 	Conversion from ProcessName to ProcessId& 	\multirow{3}*{Process} \\
  \cline{1-3} 7	&ProcessID2Name&Conversion from ProcessId to ProcessName &\\
  \cline{1-3} 8	&ProcessID2Modules&	Conversion from ProcessId to Moudle &\\
  \hline
  9	&Volume2Disk	& \makecell{File path format Conversion\\//device//harddrive $\rightarrow$ C:// }&Driver \\
  \hline
  10	&CallStackMap& Record CallStacks & \multirow{3}*{SystemCall} \\ 
  \cline{1-3} 11 & SystemCallMap&	Record SystemCall&\\  \cline{1-3} 12& SystemCallMapUsed	& Record SystemCall Used&\\
  \hline
  13	&ModulesName2APIs	&Conversion from Modules to APIs& \multirow{2}*{DLL loading} \\ 
  \cline{1-3} 14	&UsedModulesName2APIs&	Conversion from UsedModulesName to APIs &\\
  \hline
  15&	EventPropertiesMap&	Record Event Properties&\multirow{4}*{Eventstructure Config loading} \\
  \cline{1-3} 16 &	Property2IndexMap	&Conversion from Property to Index&\\
  \cline{1-3} 17&	EventStructMap& Record Operation Info and Event Attribution&\\
  \cline{1-3} 18	&Properties&	Record Properties&\\
  \hline
  \end{tabular}}
  \label{tab2}
  \end{center}
  \end{adjustwidth}
  \vspace{-1em}
  \end{sidewaystable*}

Semantic correction involves not only modifying the semantics but also updating the internally maintained data structure. This ensures that subsequent semantic modification events accurately reflect the operations of the current system. For example, the data structure related to processes needs to be updated each time a Process event is generated.

Optimizing the academic tone of the passage enhances its clarity and precision, making the content more suitable for scholarly discourse.

\textbf{Part 2: Using Callstacks to Enhance Semantic.} 

The closed nature of Windows' ETW mechanism necessitates a more in-depth analysis of its internals. we proposes an approach wherein the complete kernel log is utilized as the primary ETW provider to capture lower-level system behavior, in contrast to the limited information provided by a single behavior log. However, it should be noted that the system comprises both user and kernel sides, and not all behaviors are conveyed to the kernel through system calls. Behaviors such as keylogging, microphone monitoring, and camera monitoring predominantly rely on API executed on the user level.

To address the limitations associated with coarse-grained events that fail to capture user-side semantics, we propose an upper-layer behavior restoration method using callstacks. This technique enables the retrieval of comprehensive semantic information by capturing and recording the API sequence formed when a user implements specific functions through the user-level API. We give a more specific explanation of the callstack in the Appendix \ref{sec:CallStack}.

In security scenarios, certain attack behaviors cannot be adequately captured through coarse-grained events alone. Therefore, the combination of callstacks becomes essential in order to further identify and analyze behaviors. For instance, in the context of RAT attacks, keyboard monitoring behavior is observed. RAT achieves this functionality by registering a message callback function using the NtUserSetWindowsHookEx function. When a user interacts with the keyboard, the corresponding character virtual code is received by the callback function\cite{yang2020ratscope}. Such behavior cannot be effectively captured by relying solely on kernel logs.

It is worth noting that while the text briefly mentions Kellect's support for call stacks to enhance behavior analysis, the specific applications and implications of this approach are not elaborated upon.

\section{Evaluation}

Kellect demonstrates platform compatibility by being capable of running on Windows 7 and above releases. Furthermore, its compatibility can be extended to additional OS versions through the configuration of event structure files. In the subsequent evaluation, we will assess Kellect's performance with regards to data loss, real-time performance, and overhead.

We use the evaluation criteria mentioned in Section 2. Based on the above argument, we set the bufferisze of Kellect to 1024KB, the number of parsing threads to 5, and the buffer pool to 1024KB, and dynamically adjust the number of buffer pools according to freebuffer.

\subsection{Data Loss}

According to the previous discussion in Section 2, it has been observed that the two compared tools suffer from data loss issues in practical applications. Their limited log diversity has led us to exclude them as reference tools for comparative experiments.

In contrast, Kellect offers several distinct advantages. Firstly, our design incorporates dynamic multi-level cache adjustment to ensure no data loss occurs during event tracking. This design considers the efficient utilization and allocation of system resources, thereby minimizing the possibility of data loss and enhancing the reliability and integrity of event tracking.

Secondly, Kellect employs an efficient method for data frame parsing. By optimizing the parsing algorithm and data structure of the data frame, Kellect achieves rapid and accurate event data extraction and analysis. This efficient parsing technique enables us to capture and process large quantities of event data without experiencing data loss.

In summary, Kellect avoids data loss during event tracking and achieves reliable and comprehensive event data acquisition by implementing dynamic multi-level cache adjustment and efficient data frame parsing methods.

\subsection{Real-time Data Collection}

We employed a novel approach distinct from our previous methodology to address the real-time issue. We introduced a comparison between the time of behavior occurrence and the time when the behavior was written to the file. To accomplish this, we developed a script incorporating specific directives as flags. Since Sealighter is an open-source tool, we augmented its output statement with judgment logic without modifying its internal mechanism. If a specific command is recognized, we log the current system event and generate an output. We conducted 10 repeated tests and computed their average for subsequent analysis. The specific results are presented in table 6. Through meticulous analysis, we observed that Kellect outperformed Sealighter regarding real-time performance.

During our testing, we discovered that when subjecting Sealighter to a high-concurrency load test using JMeter, it failed to trigger the recognition of instructions even after multiple runs. We attribute this to buffer accumulation caused by the substantial volume of log data. Consequently, Sealighter's performance significantly lags behind Kellect's when processing full kernel log parsing.

\begin{table}[h]
  \caption{Real-time Test}
  \label{tab:freq}
  \begin{tabular}{p{1.6cm}p{1.6cm}p{1.6cm}p{1.6cm}}
    \toprule
     & Kellect & Sealighter & ratio \\
    \midrule
    pcopy & 0.32	&1.5	&4.69 \\
    iperf &     0.03&	0.19	&6.333 \\
    JMeter &     1.46	&N/A	&N/A \\
    rar &     0.43	&0.67	&1.56 \\
    everything &  0.34	&1.73&	5.09 \\
    IDM &    0.16	&0.42&	2.63\\
  \bottomrule
\end{tabular} 
\end{table}

\subsection{Runtime Overhead}

Through extensive benchmarking and comparative analysis against multiple tools, we have conducted a comprehensive evaluation of Kellect's performance. The results unequivocally demonstrate that Kellect outperforms other tools in terms of overall performance. As part of our evaluation, we also employed silketw as a reference tool to explore the potential advantages offered by the C\# programming language. Furthermore, we established an IDLE group as a reference baseline, which maintained minimal system operation requirements without executing any benchmark tests.

\begin{figure}[!htp] 
  \centering 
  \includegraphics[width=3.2in]{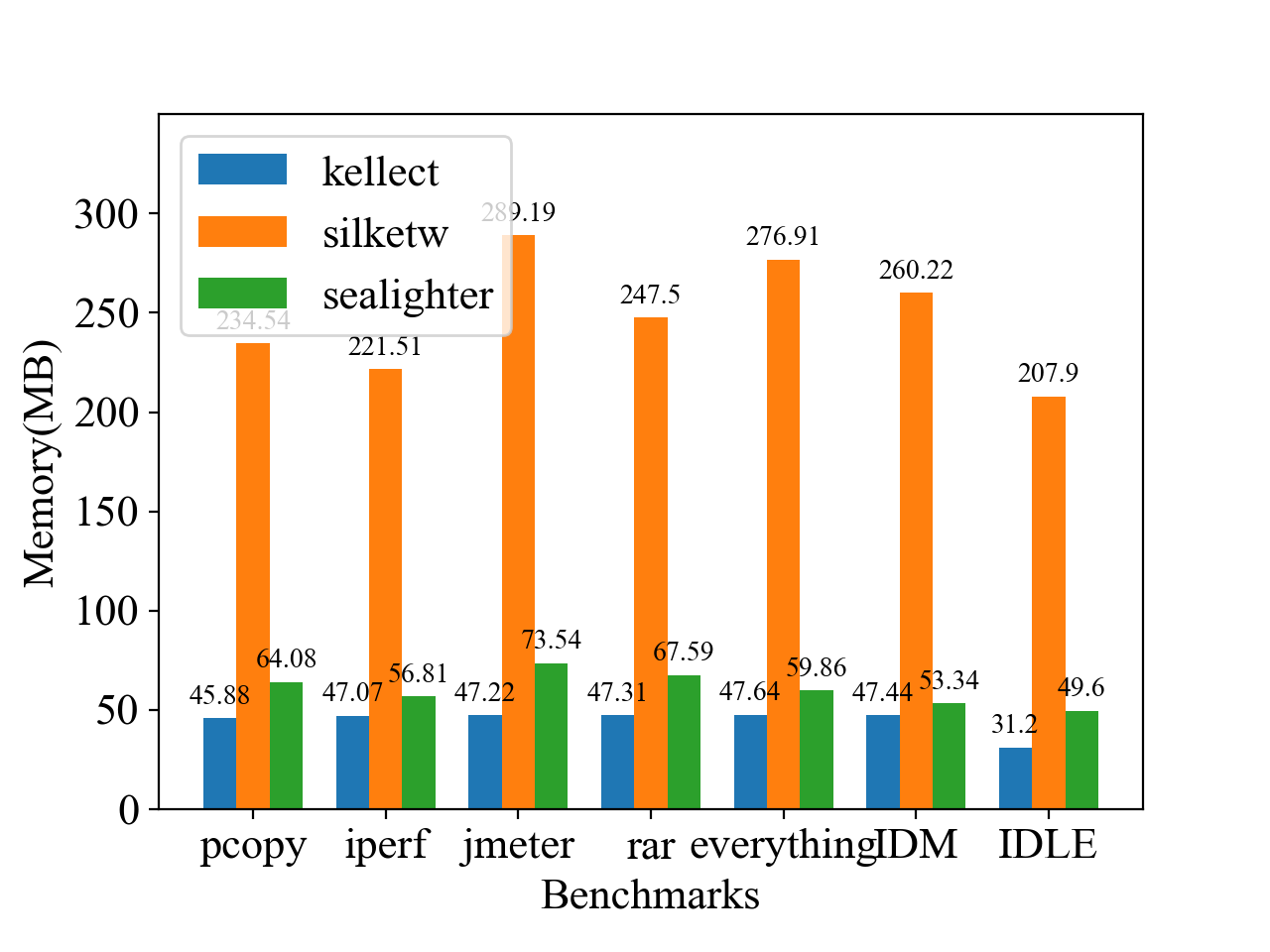}
  \caption{Memory Consumption in Different Benchmarks} 
  \label{Fig.main1} 
  \vspace{-1em}
\end{figure}

\begin{figure}[!htp] 
  \centering 
  \includegraphics[width=3.2in]{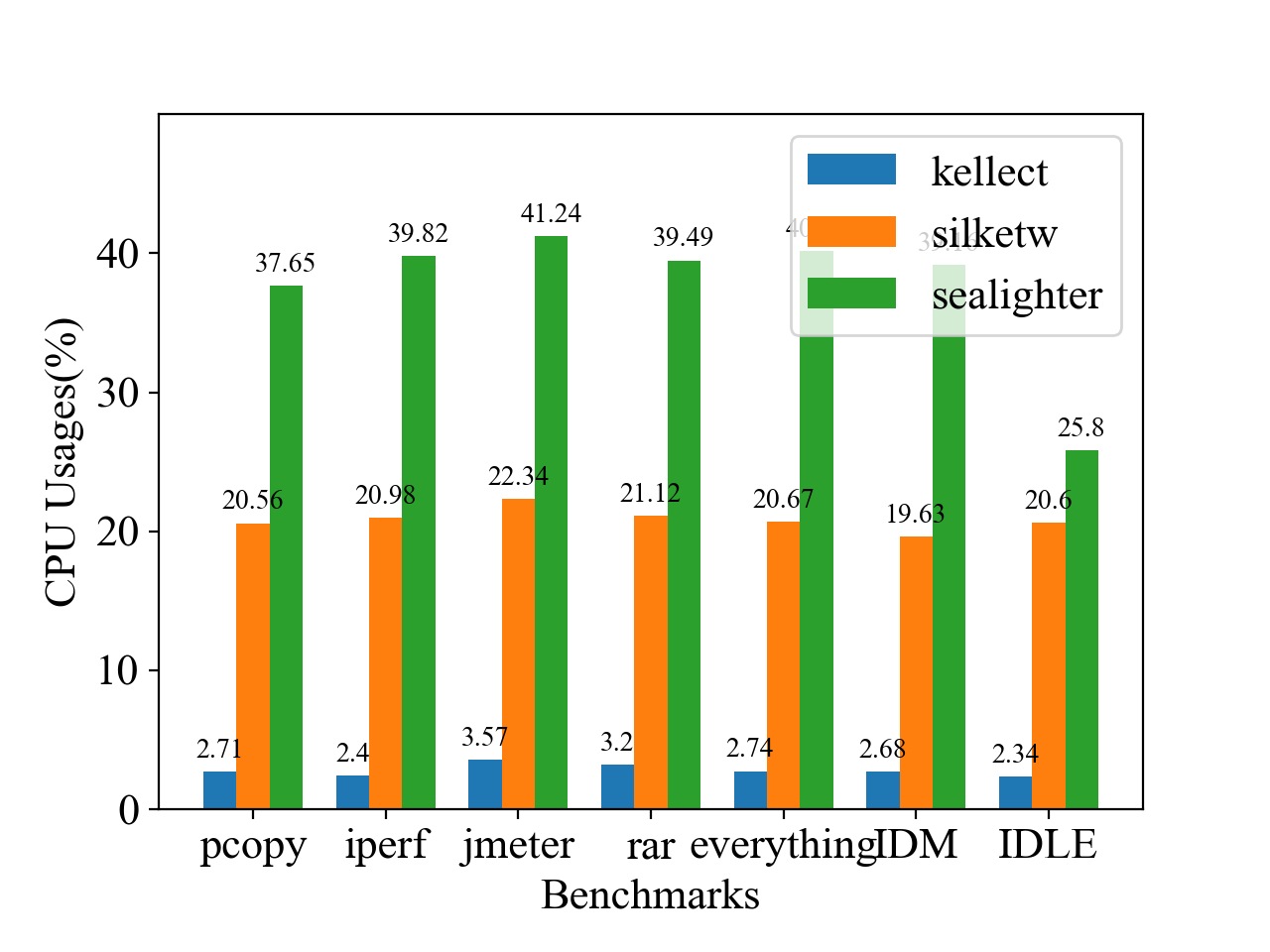}
  \caption{CPU Usage in Different Benchmarks} 
  \label{Fig.main1} 
  \vspace{-1em}
\end{figure}

Based on the findings in Figures 7 and 8, Kellect demonstrates superior overall performance compared to the other two tools. Even under full load, Kellect exhibits lower performance consumption than the IDLE group. While Silketw exhibits lower CPU usage than Sealighter, its memory consumption surpasses other tools. This discrepancy can be attributed to Silketw's slower parsing speed, leading to log accumulation in the cache and increasing memory consumption.

In the overhead experiment, we observe that as the running time increases, both Silketw and Sealighter experience a rise in CPU usage and memory consumption, which do not return to their initial values. Over time, event loss and buffer overflows exacerbate the parsing overhead of these tools. This indicates that Silketw and Sealighter fail to meet the requirements of real-time acquisition and long-term stable operation.

Regarding Kellect, its memory consumption experiences fluctuations over time due to the maintenance of multiple tables in memory for semantic correction. However, memory consumption stabilizes after the mapping table has collected complete system information. In the long-running experiment, we noted that Kellect maintains a stable CPU usage of approximately 2\%-3\% and a memory consumption of around 40MB during the no-load state.

\section{Best Practices: Kellect4APT Dataset}

As to detect real-world complex attacks, a general knowledge base from threat intelligence is always needed, depending on which a causal traceability graph derived from kernel logs can then work out to implement APT attack detection, traceability, and early warning policies \cite{milajerdi2019holmes}. In the field of network security, the ATT\&CK framework\cite{attck} has gained widespread adoption as a knowledge base and de-facto framework to study APT. It categorizes network attack techniques employed by tactical organizations, providing a robust model for supporting knowledge-based detection methods. However, relying solely on this threat intelligence may limit the identification of targeted behavior patterns. 

To address this limitation, we recognize the real-time, non-destructive, and high-efficiency capabilities of Kellect in terms of kernel log mining. Leveraging these capabilities, we utilize Kellect to collect attack data and atomic operation behavior data based on the Atomic Red Team tool proposed by Red Canary\cite{redcanaryco}. By curating a comprehensive dataset of atomic operation kernel behaviors across various technologies, our goal is to provide technique-level behavior model supporting for knowledge-driven attack detection. 

We conducted 178 ATT\&CK behaviors on Windows, representing 36 sub-techniques. In addition to carrying out the above-mentioned attack behaviors, the data collection environment also runs some normal processes, such as office activities, file transfer, software installation, and video playback. The data includes different behaviors including process, thread, image, network, fileIO, diskIO, and registry. Each piece of data contains the relationship between the subject and the object, and the detailed attributes are shown in Table 7, args represents the private attributes of specific events, such as file handles and paths, process command data, etc. The dataset describes the behavioral causality of current technologies based on ATT\&CK in Windows for the first time.

\begin{table}[h]
  \caption{Description of the Attributes}
  \label{tab:freq}
  \begin{tabular}{cl}
    \toprule
    attributes & description  \\
    \midrule
    event & the behavior's name that happened\\
tid& thread id of event\\
pid& process id of event \\
pname& process name of event\\
ppid& parent process id of event\\
ppname& parent process name of event\\
timestamp& the time that the event happend\\
host-uuid& The host ID that generated the event \\
args&Private attributes of specific events \\
  \bottomrule
\end{tabular} 
\end{table}

The atomic behaviors in Kellect4APT provide valuable knowledge support and could enhance our ability to detect and comprehend potential security threats through policy-based analysis. This policy-driven analytical approach enables precise identification and mitigation of various advanced threats, safeguarding system integrity and data security.

Thus, the recording and analysis of atomic operations with Kellect empower us to conduct in-depth research into the behavioral logic within the system, offering crucial academic support and technical capabilities for APT detection and security protection. This academic perspective optimization provides a more precise description of our research on atomic behavior, emphasizing its significance and value within the academic domain.

The effectiveness of the collected atomic operation dataset has been demonstrated in the work of APT-KGL \cite{chen2022apt}, yielding remarkable results and validating its practical utility. 

The specific details of a technique in the dataset shows in Appendix \ref{sec:Kellect4APT}. The data set has been open source.\footnote{Kellect4APT Download: \url{https://www.kellect.org}}

\section{Discussion and Future Work}

In the realm of APT detection, kernel logs have gained significant prominence as valuable data sources \cite{ahmed2021peeler,statcounter,zeng2021watson,xiong2020conan,milajerdi2019poirot}. By integrating data-driven and knowledge-driven methodologies, kernel logs can be transformed into origin graphs, facilitating causal analysis that effectively traces and alerts against APT attacks. Prominent examples of such approaches include Poirot and SLEUTH \cite{milajerdi2019poirot,hossain2017sleuth}. Given APT attacks' covert and persistent nature, conventional security defense tools often struggle to detect and respond to them on time. Our system offers enhanced APT attack detection capabilities by comprehensively collecting and analyzing kernel-level events in real time. For instance, we leverage the collected kernel log data to identify abnormal system call patterns, illicit kernel module loading, and anomalous driver behavior, enabling the discovery of potential APT attack activities.

Furthermore, our system extends its applicability to detecting other malicious behaviors. We can identify malware-related activities by analyzing kernel-level event logs, including suspicious process creation, abnormal file operations, and atypical network communications. We can bolster system security and mitigate potential threats by monitoring and promptly responding to such behaviors in real time.

\textbf{Limitations}. Firstly, the comprehensive collection of kernel logs may generate a substantial volume of data. This necessitates careful consideration of data storage and processing scalability, particularly in large-scale systems or environments with high-frequency events. Secondly, real-time requirements exert an impact on system performance and resource utilization. Thus, balancing system performance and security requirements is crucial during deployment. Although Kellect has demonstrated stable operation for extended periods in laboratory environments, the substantial amount of data generated by full kernel log collection cannot be overlooked. Therefore, efficient data compression and utilization represent important areas for future investigation. Additionally, given the extensive range of user-level providers offered by ETW, Kellect will progressively adapt and expand its capabilities to address other security issues within the system.

\textbf{Future Work}. We will further explore approaches to optimize data storage and processing to tackle the challenges posed by large-scale systems and high-frequency event environments. We will continue expanding Kellect to adapt to evolving security threats and leverage the user-level providers provided by ETW to enhance the system's security performance. Furthermore, we intend to publish the source code, test benchmarks, and the Kellect4APT dataset as open-source resources, facilitating reproducibility, further research, and practical applications by fellow researchers.

\section{Related Work}

We conducte a survey and collecte some tools and open-source projects, and divided them into three categories as table 8 shows: applications, collectors and toolkits. These tools and open-source projects aim to fill the Windows log collector gap.

\begin{table}[h]
  \caption{Tools}
  \label{tab:freq}
  \begin{tabular}{p{1.8cm}p{4.8cm}}
    \toprule
    Type &Tools\\
    \midrule
    applications & Windows Performance Toolkit\((WPT)\)\cite{performancetoolkit}, PerfView\cite{perfview}, Windows Performance Monitor\cite{windowsperformancemonitor}, UIforETW\cite{UIforETW}, Nxlog\cite{nxlog}, Wazuh\cite{wazuh} \\
    toolkits & Pywintrace\(( fireeye)\)\cite{pywintrace}, krabsetw\((microsoft)\)\cite{krabsetw}, Diagnostics\cite{Logging} \\
    collectors & Wtrace\cite{wtrace}, Spade\cite{SPADE}, Conan\cite{xiong2020conan}, Silketw\cite{SilkETW}, Sealighter\cite{Sealighter}\\
  \bottomrule
\end{tabular} 
\end{table}

Within the category of applications, several system monitoring tools rely on the ETW. Examples include WPT, PerfView, and Windows Performance Monitor provided by Microsoft, as well as UIforETW provided by Google. These tools offer a visual interface for monitoring system events, memory consumption, CPU usage, and other relevant information. However, developers face limitations in directly obtaining logs for subsequent analysis. Alternatively, monitoring systems such as Nxlog and Wazuh combine ETW with other data sources to provide a variety of terminal monitoring services.

Toolkits primarily integrate the ETW's provided APIs, delivering more convenient data collection and analysis services. For instance, FireEye offers Pywintrace based on Python, while Microsoft provides krabsetw based on C++ and Diagnostics based on C\#. Developers have utilized these toolkits to develop Windows log collectors.

In terms of collection, SilkETW is extensively employed in research areas such as attack detection and behavior analysis. Developed using C\# and relying on Diagnostics, SilkETW generates kernel logs with clear semantics. Sealighter, developed in C++ and based on krabsetw, offers enhanced analysis capabilities. On the other hand, Wtrace, developed with C\#, only supports the collection of limited data types such as files, registry, TCP/IP, and RPC calls, lacking support for all providers. Spade is a comprehensive audit log framework. While the audit data in Linux enables real-time collection, Windows requires the use of XML data generated by ProcMon (a Windows process monitoring tool)\cite{mitra2004delay} as input to generate operational behavior data, which cannot be collected in real-time. Conan, an APT detection tool published in recent years, primarily focuses on APT attack detection but is not open source.

\section{Conclusion}

This paper addresses the existing gap in the current Windows-based kernel log collection, which include data loss, poor real-time performance, and high overhead. We propose Kellect, a lossless, real-time, and efficient kernel log collection tool, to overcome these challenges. Kellect incorporates an efficient log parsing mechanism by implementing a multi-level cache scheme. It dynamically adjusts the size and number of processing threads based on the buffer of Windows ETW, effectively mitigating the issues of high overhead and data loss. Furthermore, Kellect enhances analysis efficiency by replacing the traditional TDH analysis scheme with a sliding pointer, enhancing compatibility with different OS versions. Additionally, Kellect improves log semantics by maintaining a behavior mapping table and employing the Callstack method, thus providing more comprehensive behavioral characteristics for application analysis.

With plenty of experiments, Kellect demonstrates its capability to achieve a non-destructive and real-time full collection of events with a comprehensive efficiency 9 times greater than existing tools. The log data generated by Kellect can be readily utilized in existing APT detection and source tracing methods driven by kernel logs. Moreover, leveraging Kellect and Canary technical scripts, we introduce a comprehensive behavioral dataset for ATT\&CK. This dataset serves as a valuable resource for behavior-based APT analysis.

We have released Kellect4APT dataset on the open source project website of Kellect.\footnote{Website of Kellect: \url{https://www.kellect.org}} This will enable other researchers to replicate our work, conduct further investigations, and facilitate broader research and application in this domain.

\section*{Acknowledgments}
  This work is supported partly by the following grants: Zhejiang key R\&D projects under Grant No.2021C01117; National Natural Science Foundation of China under Grant No.U22B2028 and U1936215; 2020 industrial Internet innovation and development project under Grant No.TC200H01V; Scientific research project of Zhejiang Provincial Department of Education under Grant No.Y202249647.

\bibliographystyle{sn-basic}
\bibliography{sn-article}

\clearpage
\begin{appendices}

  \begin{figure*}[htp] 
    \centering 
    \includegraphics[width=10in, angle=90]{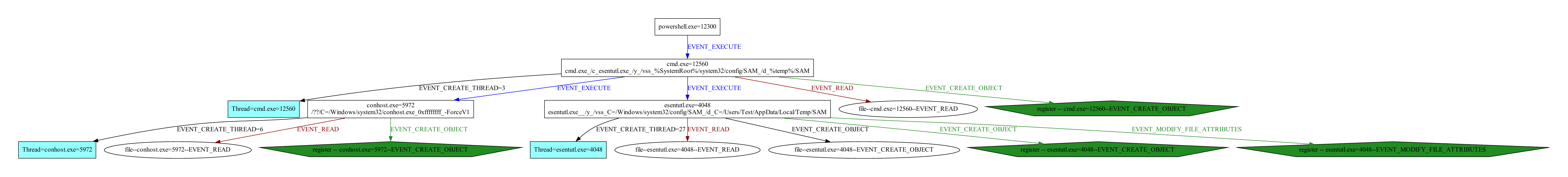}
    \caption{Provenance Graph for T1003.002-3} 
    \label{Fig.main1} 
    \vspace{-1em}
  \end{figure*}

  \section{}

  \subsection{Events That Kellect Collect}
  \label{sec:EventsCollecting}

  In our study, we have selectively collected a subset of event types that are closely associated with the operation of the Windows system. These event types include Process, Thread, Images, DiskIO, FileIO, and Registry. While Microsoft provides a wide range of related events, we did not capture all of them in our data collection process. Through our analysis, we have identified that certain events do not provide substantial insights into the behavior of the Windows system. Specifically, within the DiskIO category, we observed events named DiskIOWriteInit and DiskIOReadInit, which represent the initialization process when disk write and read operations occur. However, based on our practical analysis requirements, we determined that these two events are not relevant for our investigation and therefore were not included in our data collection efforts. Detailed information regarding the available events can be found in the official ETW documentation provided by Microsoft.
  
  \begin{table}[h]
    \caption{the Events' Types of Kellect Collects}
    \begin{tabular}{p{1.2cm}p{6cm}}
      \toprule
      Type & Event Names \\
      \midrule
      Process& ProcessStart, ProcessEnd, ProcessDCStart, ProcessDCEnd\\
      Thread& ThreadStart, ThreadEnd, ThreadDCStart, ThreadDCEnd\\
      Image& ImageDCStart, ImageLoad\\
      TCP/IP& TcpIpSendIPV4, TcpIpSendIPV6, TcpIpRetransmitIPV4, TcpIpRetransmitIPV6, TcpIpRecvIPV4, TcpIpRecvIPV6, TcpIpConnectIPV4, TcpIpReconnectIPV6, TcpIpConnectIPV6, TcpIpReconnectIPV4, TcpIpAcceptIPV4, TcpIpAcceptIPV6, TcpIpDisconnectIPV4, TcpIpDisconnectIPV6\\
      FileIO& FileIOWrite, FileIORead, FileIOFileCreate, FileIORename, FileIOCreate, FileIOCleanup, FileIOFlush, FileIOClose, FileIODelete, FileIOFileDelete\\
      DiskIO& DiskIOWrite, DiskIORead, \\
      Registry& RegistryCreate, RegistrySetValue, RegistryDeleteValue, RegistryOpen, RegistryDelete, RegistryFlush, RegistryClose, , RegistrySetInformation,  RegistryQuery, RegistryQueryValue\\
    \bottomrule
      \end{tabular}
  \end{table}
  
  \subsection{CallStack}
  \label{sec:CallStack}
  
  The process of API parsing involves obtaining specific information about an API through pointers. Specifically, we utilize the addresses from StackWalk events and previously resolved addresses to calculate the pointer values.
  
  Firstly, during the data collection process, we monitor StackWalk events to obtain the function call stack information involved in program execution. By analyzing these stack frames, we can capture the memory addresses at which API calls occur.
  
  Next, we use these memory addresses in conjunction with the previously resolved information to calculate the accurate pointer values for the APIs. This calculation process may involve adding or computing offset values to precisely locate the APIs in memory.
  
  Once we obtain the pointer values, we can accƒess the corresponding memory locations to retrieve detailed information about the APIs, including their names, parameters, return types, and other key details.
  
  Through this detailed process, we can parse the APIs and obtain their specific information, enabling subsequent semantic transformation and analysis. Ultimately, by utilizing the aforementioned parsing techniques, we can obtain the call stack of a process for conducting in-depth analysis of its specific behaviors.
  
  \subsection{Kellect4APT}
  \label{sec:Kellect4APT}
  
  Through the presented example in Figure 1, we have captured an atomic behavior using the Kellect, targeting the ATT\&CK framework. This example unveils the intricate operations taking place at the system's underlying level during script execution.
  
  Kellect, as a powerful tool, offers comprehensive insights into the specific behavioral logic of atomic operations, enabling meticulous recording. This detailed record empowers us to delve into the execution process of atomic operations within the system, including the involved resources, invoked functions, and resulting impacts.



\end{appendices}

\end{document}